\pdfoutput=1
\documentclass[a4paper,twocolumn]{autart}

\usepackage[utf8]{inputenc}
\usepackage{xcolor}
\usepackage[fleqn]{amsmath}
\usepackage{amsfonts}
\usepackage{amssymb}
\usepackage{graphicx}
\usepackage{hhline}
\usepackage{multirow}
\usepackage[percent]{overpic}
\usepackage{eucal}
\usepackage{url}
\usepackage{hyperref}

\DeclareMathOperator{\R}{\mathbb{R}}
\DeclareMathOperator{\A}{\mathcal{A}}
\DeclareMathOperator{\B}{\mathcal{B}}
\DeclareMathOperator{\C}{\mathcal{C}}

\newtheorem{theorem}{Theorem}[section]
\newtheorem{remark}[theorem]{Remark}

\newtheorem{problem}[theorem]{Problem}

\begin{document}
\begin{frontmatter}

\title{A Sum-of-Squares-Based Procedure to Approximate the Pontryagin Difference of Semi-algebraic Sets}

\author[ULB]{Andres Cotorruelo\corauthref{Corres}}\ead{acotorru@ulb.ac.be},
\author[MI]{Ilya Kolmanovsky}\ead{ilya@umich.edu},
\author[ULB]{Emanuele Garone}\ead{egarone@ulb.ac.be}

\corauth[Corres]{Corresponding author.}

\address[ULB]{Service d'Automatique et d'Analyse des Syst\`{e}mes (SAAS), Universit\'{e} Libre de Bruxelles (ULB), Brussels, Belgium}
\address[MI]{Department of Aerospace Engineering, University of Michigan, MI, U.S.A.}


\maketitle

\thanks{Special thanks to Philip Pugeau for providing some of the case studies used in this paper. The second author  acknowledges the support of the National Science Foundation grant 1931738.}
\begin{abstract}
The P-difference between two sets $\mathcal{A}$ and 
$\mathcal{B}$ is the set of all points, $\mathcal{C}$, such that the addition of $\mathcal{B}$ to any of the points in $\mathcal{C}$ is contained in $\mathcal{A}$. Such a set difference plays an important role in robust model predictive control and in set-theoretic control.
In this paper we demonstrate that an inner approximation of the P-difference between two semi-algebraic sets can be computed using the Sums of Squares Programming, and we illustrate the procedure using several computational examples. 
\end{abstract}
\end{frontmatter}

\section{Introduction}
The Pontryagin set difference (or simply P-difference), so named after Pontryagin who used it in the setting of game theory \cite{pontryagin1967linear},  has become an indispensable part of robust model predictive control (MPC) \cite{kouvaritakis2016model,rawlings2017model} and of set theoretic control \cite{blanchini2008set,kolmanovsky1998theory}.  
Additionally, the P-difference has also been used in image processing applications~ \cite{heijmans1995mathematical} and in path planning \cite{luo2018porca}. 
Depending on the authors, the P-difference is sometimes referred to in the literature as a Minkowski set difference or as a set erosion.
Efficient procedures to compute the P-difference or its approximations, especially in the case of non-polyhedral sets, can greatly expand the range of applications of robust MPC.

In this paper, we demonstrate that an inner approximation of the P-difference between two semi-algebraic sets can be computed using Sum of Squares Programming (SOSP).  Computational examples are reported to illustrate the proposed approach.

\subsection*{Notation}
The ring of polynomials in the variables $x_1,\ldots,x_n$ and with coefficients in the field $\R$ is denoted by $\R[x_1,\ldots,x_n]$. The set of all Sum of Squares polynomials in the variables $x_1,\ldots,x_n$ is denoted by $\Sigma[x_1,\ldots,x_n]$. For a number of elements $a_1,\ldots,a_n$, $\{a_i\}_{i=1}^n$ denotes the set $\{a_1,\ldots,a_n\}$, and every operator applied to it is meant to be understood element-wise, \textit{e.g.} $\{a_i\}_{i=1}^n\geq 0$ means $a_i\geq0,~\forall i=1,\ldots,n$.
\section{Problem statement}
\begin{figure}
    \centering
    \begin{overpic}[trim={1.2cm 2cm 1.2cm .5cm},clip,width=.75\linewidth]{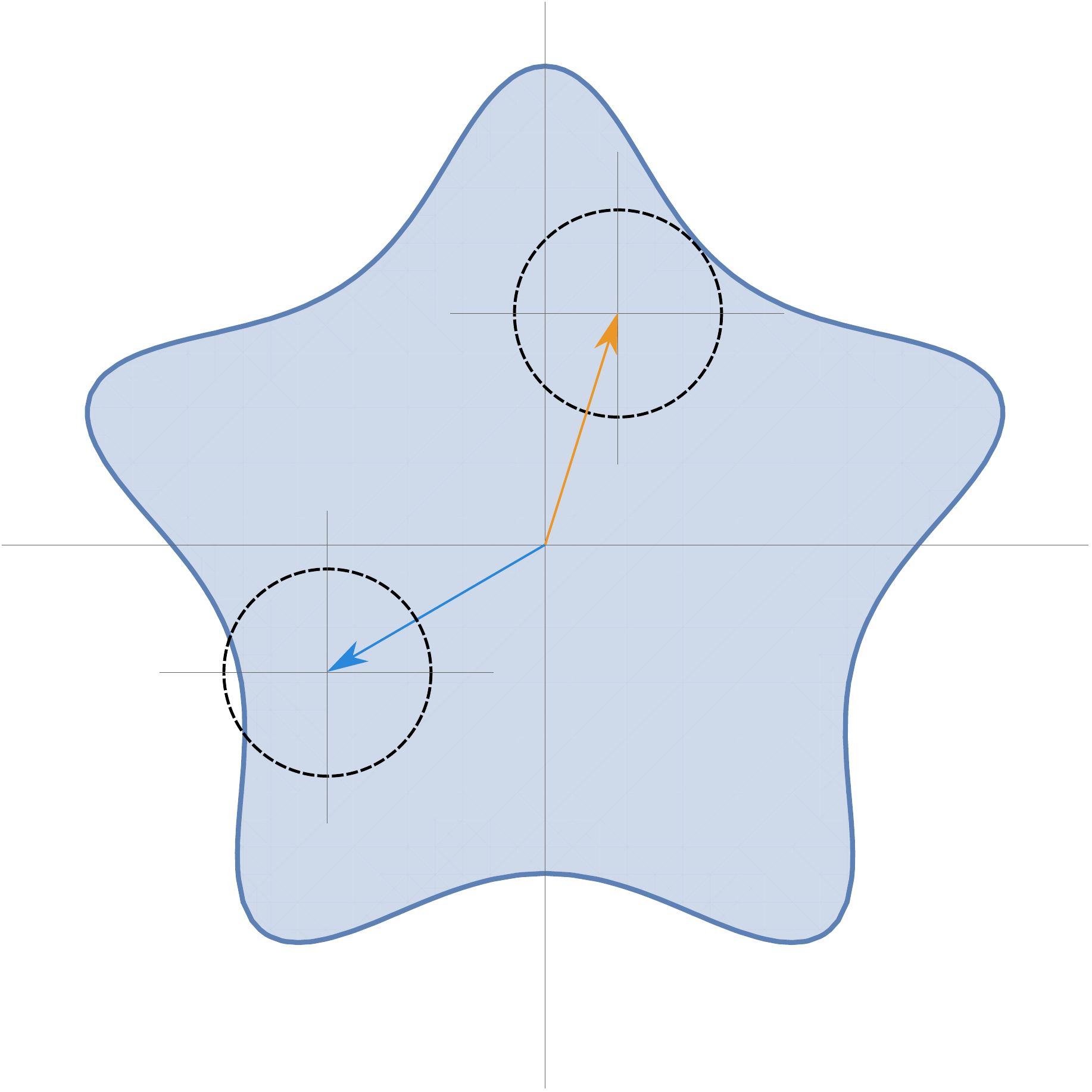}
    \put (47,55) {$x_1$}
    \put (40,35) {$x_2$}
    %\put (62,73) {$w$}
    \put (10,55) {\large$\A$}
    %\put (72,75) {\large$\B$}
    \end{overpic}
    \caption{Visualization of the Pontryagin difference between sets $\A$, the blue star, and $\B$, the dashed-line ball. As it can be seen, $x_1$ will belong to $\A \ominus \B$ since every vector $z \in \B$ is such that $x_1+z\in\A$; on the contrary, $x_2$ does not belong to $\A \ominus \B$ since there exist some $z'\in\B$ such that $x_2+z'\notin \A$.}
    \label{fig:visualization}
\end{figure}

For two sets $\A\subset \R^n$ and $\B\subset\R^n$, the %Pontryagin--Minkowski set difference (P-difference) 
P-difference is defined as
\begin{equation*}
    \A \ominus \B=\{x\in \A : x+z\in \A~\forall z\in \B\},
\end{equation*}
where typically $0\in \B$. A simple geometrical way to interpret this operation is that the set $\mathcal{C}=\A \ominus \B$ is a set so that if we select a  point in $\mathcal{C}$ and we add an uncertainty bounded by $\B$, the resulting point still belongs to $\A$ (see \figurename~\ref{fig:visualization}).

Although some algorithms able to perform this operation in  the case of polyhedral $\A$ and convex $\B$ \cite{althoff2015computing,barki2010new} exist in the literature, to the best of the authors' knowledge there exists no systematic way to compute the Pontryagin difference for wider classes of sets. The objective of this paper is to solve the following problem
\begin{problem}{(Pontryagin difference)}\label{pro:1}
Let $\A$ and $\B$ be two semi-algebraic sets in the form
\begin{equation*}
\begin{aligned}
  \A&=\{x:a_i(x)\geq 0,\,i=1,\ldots,n\},\\
  \B&=\{x:b_j(x)\geq 0,\,j=1,\ldots,m\},
  \end{aligned}
\end{equation*}
where $a_i(x),b_j(x) \in \mathbb{R}[x],~i=1,\ldots,n, ~j=1,\ldots,m$. Determine a systematic procedure to compute an inner approximation of $\A \ominus \B$.
\end{problem}
\section{Computation of the Pontryagin Difference using Sum of Squares}
In this section we  propose a way to solve Problem~\ref{pro:1} for the case where  the sets $\A$ and $\B$ are described as the intersection of polynomial inequalities. The proposed solution makes use of the  Krivine -- Stengle Positivstellensatz (P-satz) \cite{stengle1974nullstellensatz}.

To simplify the problem, the first step is to note that the set $\A$ can be represented as
\begin{equation*}
    \A=\bigcap_{i=1}^n \A_i
\end{equation*}
where $\A_i=\{x:a_i(x)\geq 0\},\,i=1,\ldots,n$. Since
\begin{equation*}
    \A \ominus \B = \bigcap_{i=1}^n \A_i \ominus \B,
\end{equation*}
we can focus on a single set $\A_i$ at a time without any loss of generality. Consider the P-difference  $    \A_i \ominus \B$
\begin{equation*}
    \A_i \ominus \B = \{x:a_i(x+z)\geq 0,\,z\in\B\}.
\end{equation*}
A possible way to approximate $\A_i \ominus \B$ is by means of a set $
    \C_i = \{x:c_i(x)\geq 0\}\subseteq\A_i \ominus \B,
$
where the function $c_i$ must be such that
\begin{equation}\label{eq:fdelta}
    c_i(x)\geq \min_{z \in\B} a_i(x+z).
\end{equation}
Note that whenever \eqref{eq:fdelta} is an equality, $\C_i = \A_i \ominus \B$.

Condition \eqref{eq:fdelta}  is equivalent to the following set emptiness condition
\begin{equation}\label{eq:emptyset0}
    \left\{ x,z: c_i(x) - a_i(x+z)<0,\, z \in \B \right\} = \emptyset.
\end{equation}
Since in the Krivine--Stengle P-satz, the set required to be empty is described in terms of \textit{equal-to}, \textit{greater-than-or-equal-to}, and \textit{not-equal-to} operators, the set \eqref{eq:emptyset0} is rewritten in terms of these operators as 
\begin{multline}\label{setempty}
    \Big\{x,z:c_i(x) - a_i(x+z) \geq 0,\\ c_i(x) - a_i(x+z) \neq 0, \, \left\{b_j(z)\right\}_{j=1}^m \geq 0 \Big\}= \emptyset.
\end{multline}
At this point the Krivine--Stengle P-satz states that \eqref{setempty} is satisfied if and only if there exist two polynomials $p(x,z)$ and $q(x,z)$ such that
\begin{equation*}
    p(x,z)+q^2(x,z)=0,
\end{equation*}
where
\begin{equation*}\begin{aligned}
    p & \in \textnormal{Cone} \left( \left\{ c_i(x) - a_i(x+z), b_1(z), \ldots ,b_m(z) \right\} \right),\\
    q&\in\textnormal{Monoid}\left(c_i(x)-a_i(x+z)\right).
    \end{aligned}
\end{equation*}
Performing standard algebraic manipulations this allows to obtain the sufficient condition
\begin{multline*}
    \left(c_i(x) - a_i(x+z)\right)^2 + s_0(x,z) \left(c_i(x) - a_i(x+z)\right)\\
    + \sum_{j=1}^m s_j(x,z) b_j(z) \left(c_i(x) - a_i(x+z)\right)=0
\end{multline*}
where $s_i(x,z)\in\Sigma[x,z]$, $i=0,\ldots,m$. This equation can be further simplified by cancelling  $c_i(x)-a_i(x+z)$
\begin{equation*}
    c_i(x)-a_i(x+z) + s_0(x,z) + \sum_{j=1}^m s_j(x,z) b_j(z) = 0.
\end{equation*}

Since $s_0\in\Sigma[x,z]$, it follows that the latter is equivalent to 
\begin{multline}\label{eq:SOSfinal}
P_i(x,z)=a_i(x+z)-c_i(x)\\-\sum_{j=1}^m s_j(x,z) b_j(z)\in\Sigma[x,z].
\end{multline}

Finally, since we are interested in the largest inner approximation of $\A \ominus \B$, using  \eqref{eq:SOSfinal} we can define the problem of finding $c_i(x)$
as the
following Sum of Squares Programming (SOSP) problem
\begin{equation}\label{eq:opt}
\begin{array}{lrl}
      \max {\displaystyle \int_\mathcal{R}  c_i(x)}~\textnormal{d}x&&\\
\text{s.t.} \quad &P_i(x,z)&\in\Sigma[x,z]\\
           &\{s_j(x,z)\}_{j=1}^m&\in\Sigma[x,z],
\end{array}
\end{equation}
where $\mathcal{R}$ is a normal domain \cite{calculus} that contains $\A$. As is well known \cite{parrilo2000structured}, optimization problem \eqref{eq:opt} can in turn be cast into a Semi-Definite Programming (SDP) optimization problem that can be solved efficiently using existing SDP solvers.

\begin{remark}
Note that whenever $\mathcal{R}$ is a normal domain described by polynomials, $\int_\mathcal{R} c_i(x) \textnormal{d}x$ can  be computed in closed form and is polynomial \cite{algebra}, which implies that the objective function of \eqref{eq:opt} is linear in the coefficients of $c_i(x)$. If one prefers to not use a normal set $\mathcal{R}$, a practical approach is to randomly select a (possibly large) number of points $r_1,\ldots,r_N \in \mathcal{R} $, and use the objective function
\begin{equation}\label{eq:objfun2}
    \frac{1}{N}\sum_{j=1}^N c_i(r_j).
\end{equation}
Note that for a sufficiently large $N$, optimizing over \eqref{eq:objfun2} is equivalent to optimizing over $\int_\mathcal{R}c_i(x)~\textnormal{d}x$. 
\end{remark}

%
%\begin{equation}\label{eq:opt_sdp}
%\begin{array}{lrl}
%  {\displaystyle \max_{\kappa_i,\{\sigma_j\}_{j=1}^m}} ~w^T\kappa_i&&\\
%\text{s.t.} \quad &\Gamma_i&\geq 0\\           %&\{\Gamma_{s_j}\}_{j=1}^m&\geq 0 ,
%\end{array}
%\end{equation}
%
%where, for a fixed $i$, $\kappa_i$ and $\sigma_j$ are respectively the vector of coefficients of $c_i(x)$ and $s_j(x,z)$ in the chosen basis, $w$ is the vector of weights of $\kappa_i$ as a result of integrating $c_i(x)$ over $\mathcal{R}$, and $\Gamma_i$ and $\Gamma_{s_j}$ are respectively the Gram matrices of polynomials $P_i(x,z)$ and $s_j(x,z)$ in the chosen basis.
%
%
\section{Examples}
\begin{figure}
    \centering
    \includegraphics[width=.8\linewidth]{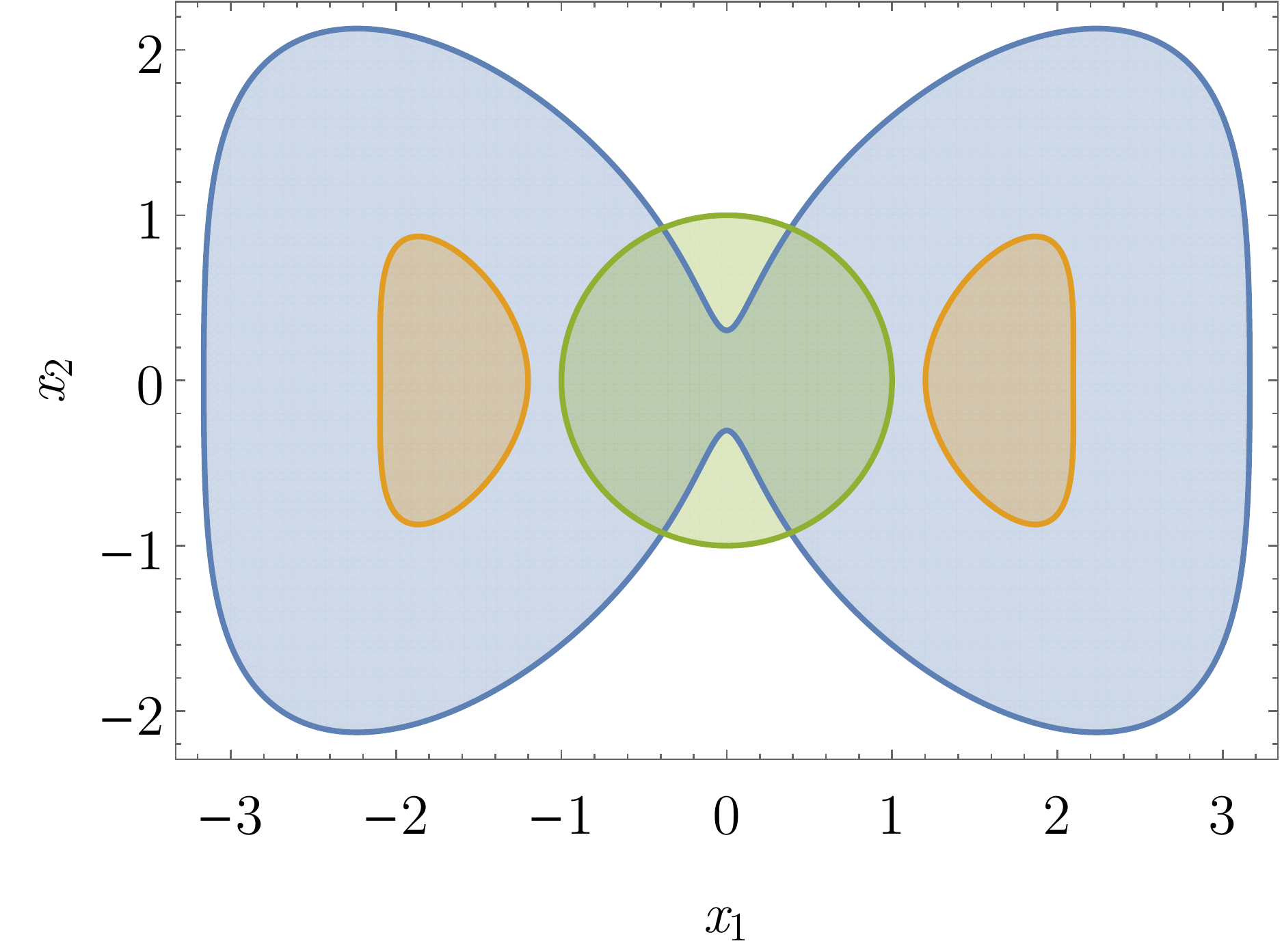}
    \caption{Result of subtracting the norm-2 ball to the \textit{bow-tie} set.}
    \label{fig:first}
\end{figure}

\begin{figure}
    \centering
    \includegraphics[width=.8\linewidth]{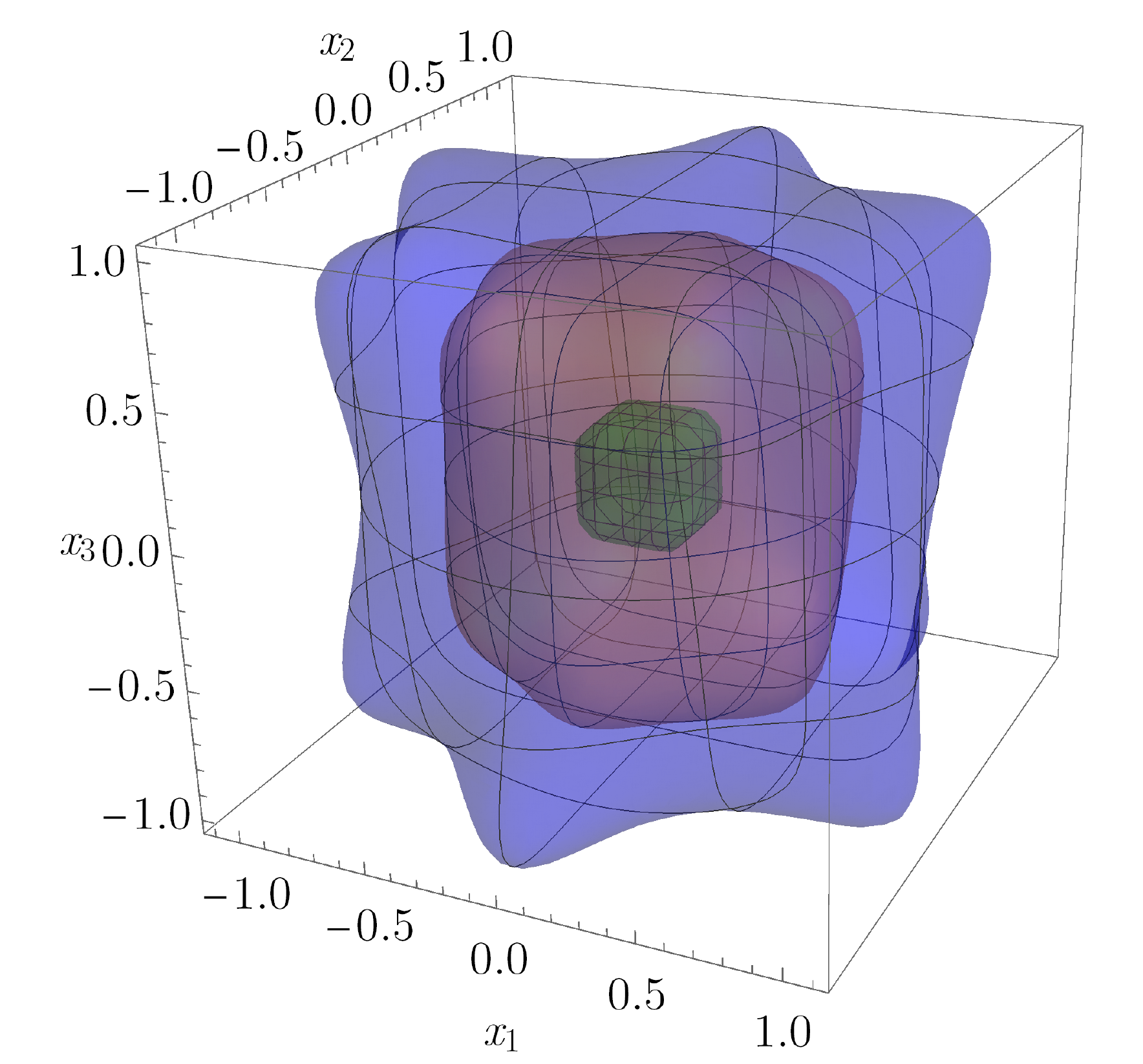}
    \caption{Result of subtracting the 6-norm ball from the rotated 5-pointed star algebraic cylinder.}
    \label{fig:last}
\end{figure}

In this section we apply the proposed methodology to a number of 2 and 3-dimensional sets to illustrate its effectiveness. All of the showcased examples depict $\mathcal{C} \approx \A \ominus \B$, where $\A= \{x:a(x)\geq 0\}$, and $\B=\{x:b(x)\geq 0\}$ with varying $a(x)$ and $b(x)$ depending on the example. Table \ref{tab:my_label} reports the expressions of $a(x)$ and $b(x)$ as well as the chosen degrees of $c(x)$ and the $s_i(x,z)$, and the elapsed time to compute the approximation. Lastly, Figs. \ref{fig:first}--\ref{fig:last} depict $\A$ as a solid blue set, $\B$ as a solid green set, and $\mathcal{C}$ as a solid orange set. For space reasons, the expressions of $c(x)$ have been omitted in this paper, but they can be found in the addendum \url{http://www.gprix.it/SoSPontryagin.pdf} . All optimization problems were solved using MATLAB R2019b and YALMIP \cite{Lofberg2004}, running on an Intel Core i7-7500 at 2.7 GHz with 16 GB of RAM.

\begin{table*}
    \centering
    \begin{tabular}{|c|c|c|c|c|c|}
        \hline Fig. &$a(x)$ & $b(x)$ &  $\partial c(x)$ & $\partial s_j(x,z)$ & $t$ \\ \hhline {|=|=|=|=|=|=|}
         2& $0.1-x_1^4-x_2^4+10x_1^2-x_2^2$ & $1-x_1^2-x_2^2$ & 14 & 6 & 11.93 s\\ \hline
         3 & $x_2^4-(x_1 - 0.5)^3 - (x_1 - 0.5)^4$ & $0.1-2x_1^2-16x_2^2$ & 10 & 6 & 2.63 s\\ \hline
         \multirow{2}{*}{4} & \multirow{2}{*}{$4-x_1^2-x_2^2$} & $0.1-25x_1^2 x_2^2$ & \multirow{2}{*}{10} & \multirow{2}{*}{2}& \multirow{2}{*}{1.02 s}\\
         &&$ - 0.05(x_1 + x_2)^2$&&&\\\hline
         \multirow{2}{*}{5} & $-(x_1^2+x_2^2+x_3^2)^3 +3(x_1^2+x_2^2+x_3^2)^2$ & \multirow{2}{*}{$0.1 - x_1^2-x_2^2-4x_3^2$} & \multirow{2}{*}{10} & \multirow{2}{*}{4} & \multirow{2}{*}{52.11 s}\\
         &$-9(x_1^2+x_2^2+3) +16(x_1^3-3x_1x_2^2+2x_3^2)$&&&&
         \\\hline
         \multirow{2}{*}{6} & $1- x_1^6 - x_2^6 - x_3^6 +5x_1^4x_2x_3 - 3x_1^4x_2^2$ & \multirow{2}{*}{$10^{-4}-x_1^6-x_2^6-x_3^6$} &\multirow{2}{*}{10} & \multirow{2}{*}{4} & \multirow{2}{*}{29 min}\\
         &$- 10x_1^2x_2^3x_3 - 3x_1^2x_2^4 + x_2^5x_3$&&&& \\ \hline
    \end{tabular}
    \caption{Polynomials used in the Figures.}%. All of them are of the form $A\ominus B$ (depicted in orange), with $A=\{x:a(x)\geq0\}$ (blue), and $B=\{x:b(x)\geq 0\}$ (green).}
    \label{tab:my_label}
\end{table*}

\begin{figure}
    \centering
    \includegraphics[width=.6\linewidth]{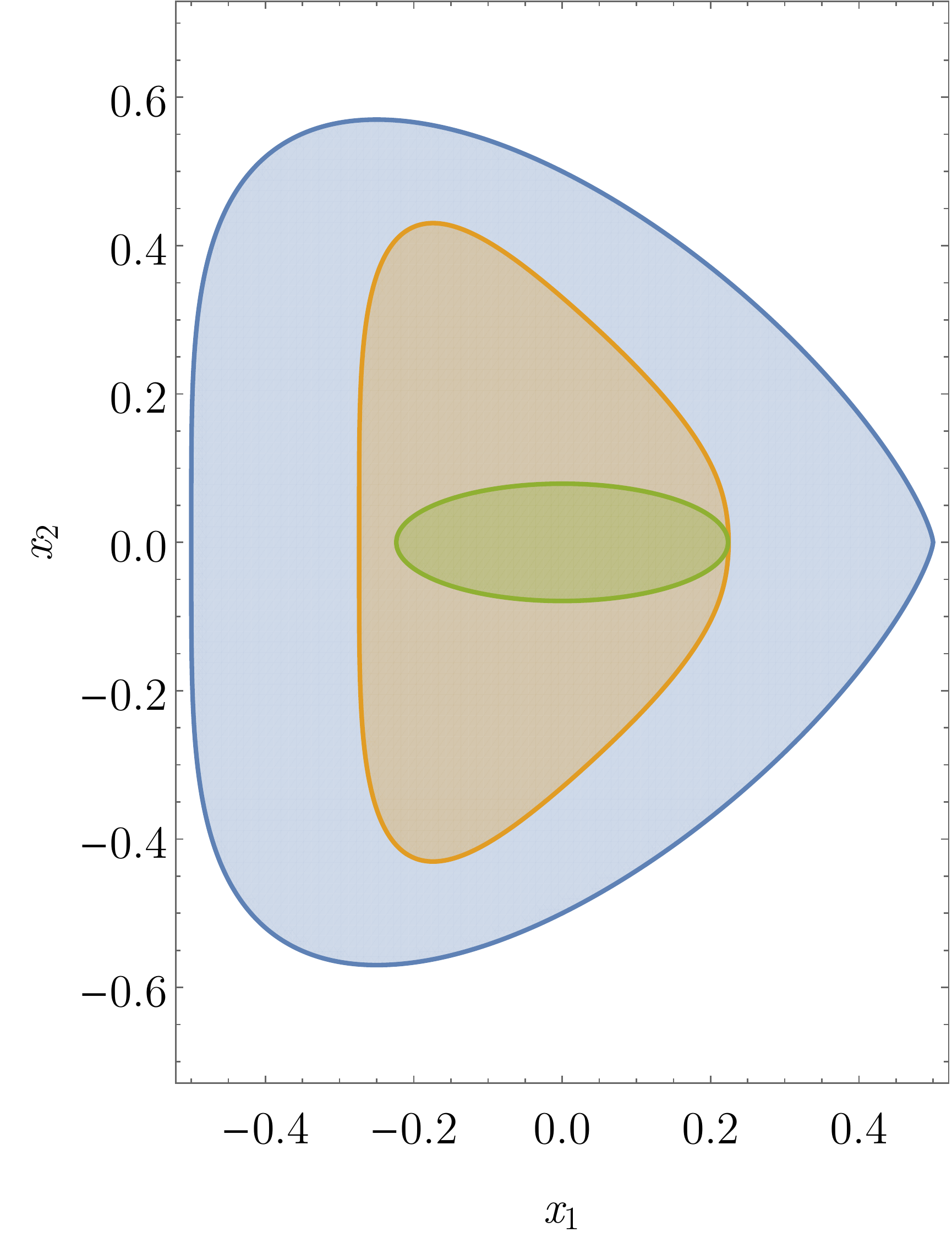}
    \caption{Result of subtracting an ellipsoid to the \textit{guitar pick} set.}
\end{figure}

\begin{figure}
    \centering
    \includegraphics[width=.6\linewidth]{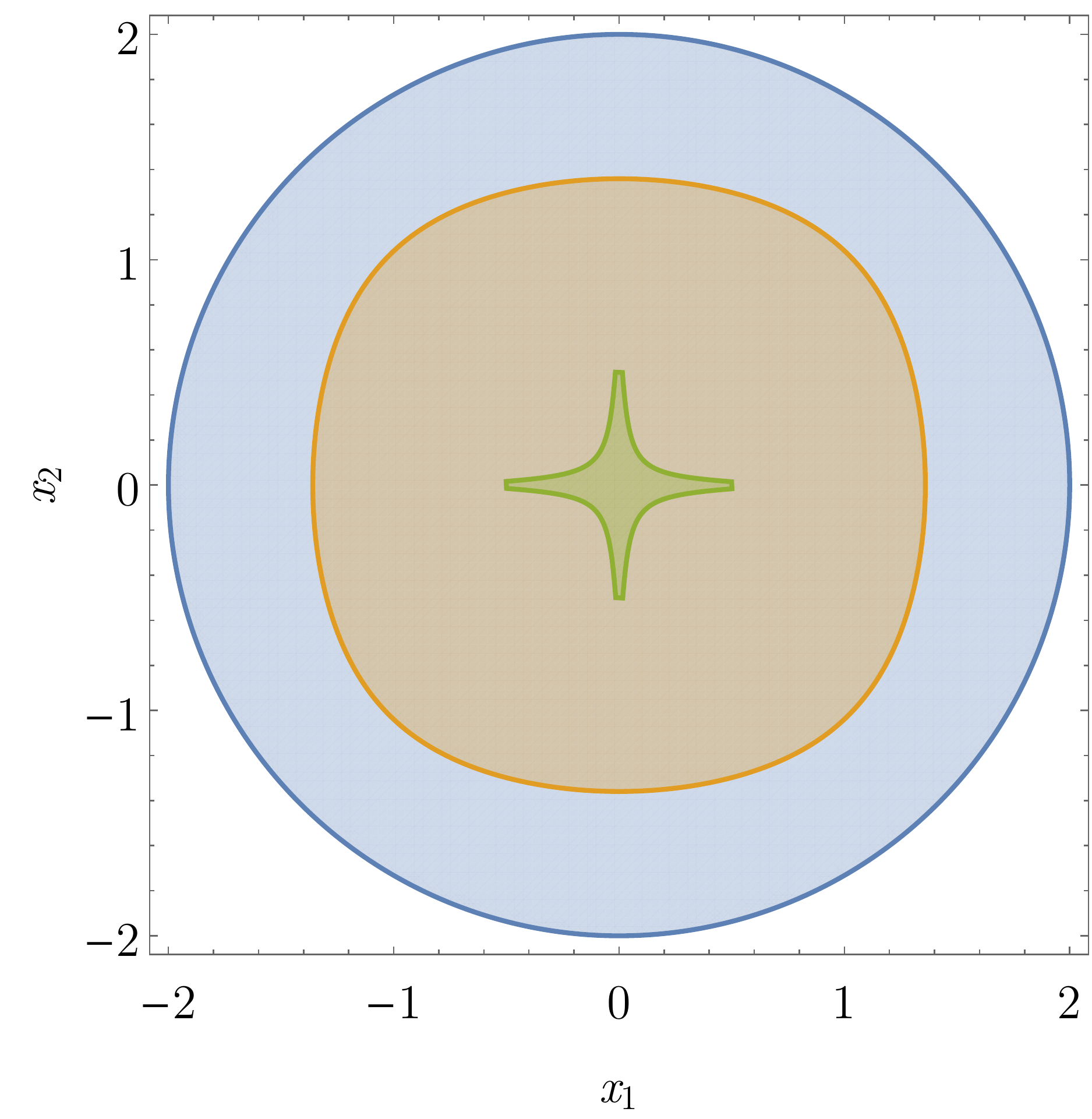}
    \caption{Result of subtracting a 4-pointed star-shaped set from the norm-2 ball.}
\end{figure}

\begin{figure}
    \centering
    \includegraphics[width=.7\linewidth]{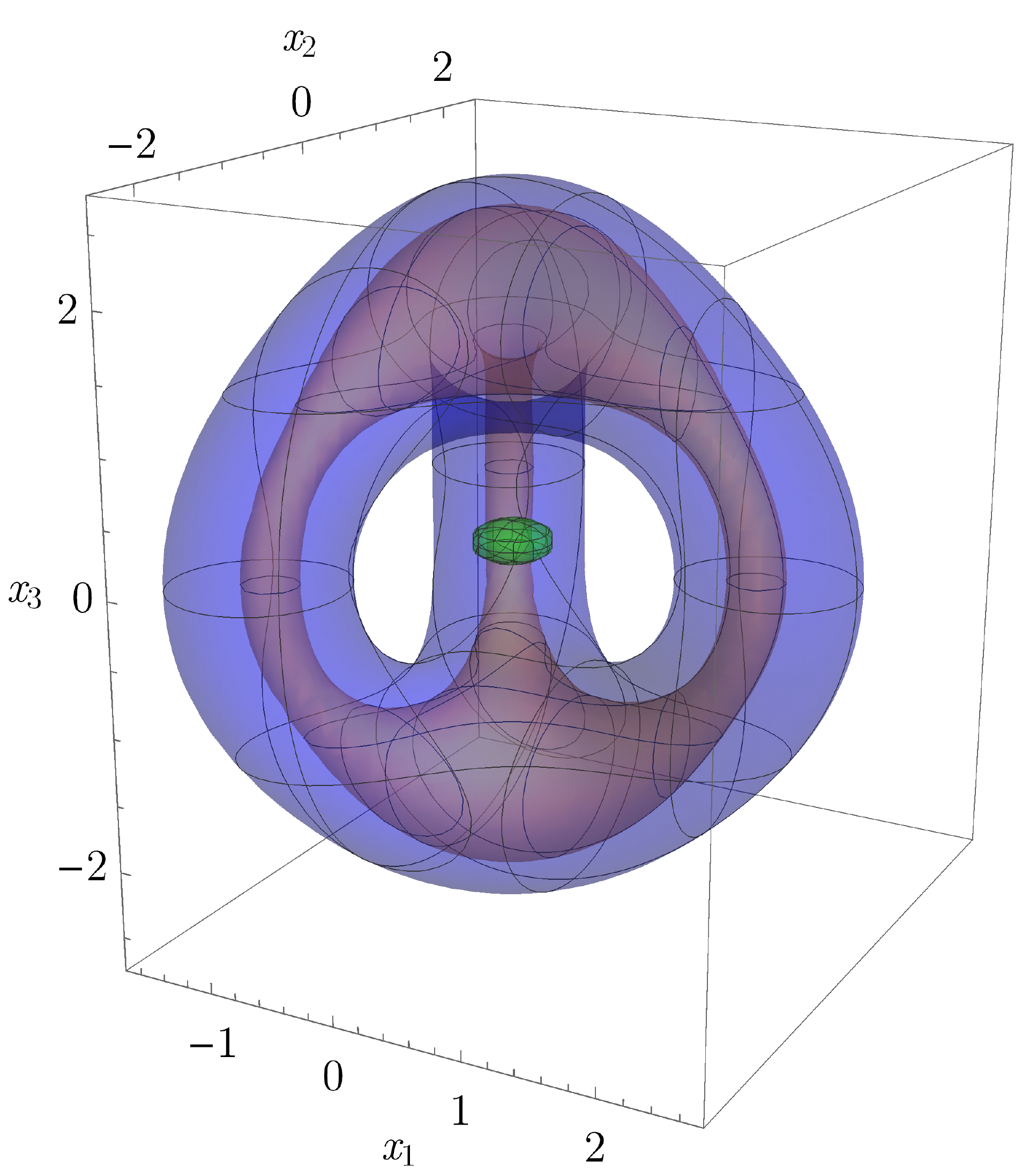}
    \caption{Result of subtracting an ellipsoid from the 3-dimensional 2-torus.}
\end{figure}

\section{Concluding remarks}
In this paper we proposed a systematic approach based on SOSP for the computation of an inner approximation of the Pontryagin difference between two semi-algebraic sets. We subsequently showcased the capabilities of this methodology by applying it to several different examples in two and three dimensions. Possible applications for this methodology include the analytical determination of an inner approximation of constrained sets in robust control.

\bibliography{bib.bib}

\end{document}